\documentclass[preprintnumbers,amsmath,amssymb,floatfix,11pt,prd,onecolumn, showpacs, showkeys,superscriptaddress,nofootinbib]{revtex4}
\usepackage{latexsym,float}
\usepackage{epsfig}
\usepackage{epstopdf}
\usepackage{amssymb}
\usepackage{verbatim}
\usepackage{multirow}
\usepackage{color}
\usepackage{tikz}
\usetikzlibrary{matrix}
\begin{document}
\title{\bf Noether Symmetry Approach in $f(T,B)$ teleparallel cosmology}

\author{Sebastian Bahamonde}
\email{sebastian.beltran.14@ucl.ac.uk}\affiliation{Department of Mathematics, University College London,
    Gower Street, London, WC1E 6BT, UK,}
\author{Salvatore Capozziello}
\email{capozziello@na.infn.it}
\affiliation{Dipartimento di Fisica, Universit\'a di Napoli
	\textquotedblleft{Federico II}\textquotedblright, Napoli, Italy,}
\affiliation{Gran Sasso Science Institute, Via F. Crispi 7, I-67100, L' Aquila,
	Italy,}
\affiliation{INFN Sez. di Napoli, Compl. Univ. di Monte S. Angelo, Edificio G, Via
	Cinthia, I-80126,
	Napoli, Italy.}

\date{\today}

\begin{abstract}
We consider the cosmology derived from   $f(T,B)$ gravity where $T$ is the torsion scalar  and $B=\frac{2}{e}\partial_{\mu}(e T^{\mu})$  a boundary term. In particular we discuss how it is possible to recover, under the same standard,   the teleparallel $f(T)$ gravity,  the curvature $f(R)$ gravity and the teleparallel-curvature $f(R,T)$ gravity, which are particular cases of $f(T,B)$. We adopt the Noether  Symmetry Approach to study the related dynamical systems and to find out cosmological solutions.
\end{abstract}

\pacs{98.80.-k, 95.35.+d, 95.36.+x}
\keywords{Modified gravity;  cosmology; Noether symmetries;  exact solutions.} 

\maketitle

\section{Introduction}

Nowadays, one of the most important problems in Physics is to understand the late-time accelerated expansion of the universe. 
Besides, large scale structure, ranging from galaxies to superclusters, presents the problem of missing matter, i.e. the luminous matter is not sufficient in order to guarantee the stability and the evolution of self-gravitating astrophysical systems.
There are several candidates to explain these phenomena and  the most popular ones are the dark energy and the dark matter, i.e. cosmic fluids gravitationally interacting and leading the evolution of the Hubble flow but without any electromagnetic counterpart.   

In particular, the physics underlying the dark energy is still not understood since it behaves as a repulsive gravitational force in contrast to the usual gravitational field. In general, there exists two ways to study the dark energy problem: i) one can retain the General Relativity (GR) and introduce a new kind of fluid which possess a negative pressure (e.g. by introducing a scalar field), or ii) one can think that GR needs to be modified at high energy levels and hence the dark energy comes out from these modifications.  One  can change the left hand side of the Einstein field equations related with the theory of gravity (e.g. extending GR or considering alternatives \cite{Mauro}) or one can modify the right hand side of it by changing the matter content of the universe (i.e., the energy-momentum tensor).

Beside the issue to explain the energy-matter content of the universe, and then the source of accelerated expansion and structure aggregation, competing theories of gravity are posing several fundamental questions on the nature of gravitational field. 
In particular, if torsion has to be involved in dynamics, if equivalence principle is valid in any case, if geodesic structure and metric structure are related or not, if theories of gravity have to be formulated in metric, metric-affine or purely affine approaches \cite{rep1, rep2}. In particular, the teleparallel formulation of gravity is recently acquiring a lot of interest due to its applications at cosmological and fundamental level. 

 In this paper, we are interested on studying cosmology in teleparallel modified theories of gravity, which in contrast to GR, consider a curvatureless space-time with a non-zero torsion. In this perspective, one needs to introduce the so-called Weitzenb\"ock connection instead of  the standard  Levi-Civita connection \cite{Weit}. By doing so, a space-time endorsed with non-zero torsion and a vanished curvature is achieved. 

From the geometrical point of view, these space-times are different than the ones considered in GR. From the point of view of the field equations, the teleparallel equivalent of General Relativity (TEGR) is  equivalent  to GR (see \cite{Aldrovandi:2013wha,Maluf:2013gaa} for further notions of TEGR). A natural extension of TEGR is, instead of considering only the trace of the torsion tensor  $T$ in the action,  to introduce a function $f(T)$ in it (see the review paper \cite{Cai:2015emx} for a discussion and references therein). $f(T)$ gravity remains  a second order theory whereas the straightforward GR extension, $f(R)$ gravity is a fourth-order one in metric formalism.  Although, the price to pay in this approach is that $f(T)$ is not invariant under local Lorentz transformations, and hence different vierbein could arise different field equations \cite{Sotiriou:2010mv,Li:2010cg}. This theory has been used in cosmology to understand the accelerating cosmic expansion of the universe  \cite{Ferraro:2006jd,Bengochea:2008gz,Linder:2010py,Saridakis1}, reconstruct cosmological models based on observational data \cite{Wu:2010mn,Bengochea:2010sg,Saridakis2,Saridakis3}, among other studied.  

Recently, a new generalisation of the standard $f(T)$ gravity was proposed in \cite{Bahamonde:2015zma}. In this theory, the function $f(T)$ is extended to $f(T,B)$, where $B$ is a boundary term related with the Ricci scalar via $R=-T+B$. By adding this dependency, one can recover $f(T)$ and $f(R)$ under suitable limits. Some cosmological features as reconstruction techniques and thermodynamics have been  studied in \cite{Bahamonde:2016cul} under the standard of this new  theory. In \cite{Bahamonde:2015hza}, a non-minimally coupled scalar field with both the boundary term and the torsion scalar was presented in view to study cosmology by dynamical system techniques. There, it was shown that a dynamical crossing of the phantom barrier is possible and also without fine tuning, the system evolves to a  late-time acceleration attractor solution. Some exact solutions and its thermodynamics properties were also discussed in \cite{Zubair:2016uhx}. In summary $f(T,B)$ gravity present several interesting features by which it is possible to unify, under the same standard, issues coming from $f(T)$ and $f(R)$ gravity.

Here, we will explore cosmological solutions coming from $f(T,B)$ by the so-called {\it Noether Symmetry Approach}.
This technique proved to be very useful for  several reasons: $i)$ it  allows to fix physically interesting  cosmological models related to  the conserved quantities (i.e. in particular couplings and potentials) \cite{cimento}; $ii)$ the existence of Noether symmetries allows to reduce dynamics and then to achieve exact solutions \cite{NoetherSergei}; $iii)$ symmetries act as a sort of selection rules to obtain viable models in quantum cosmology \cite{lambiase}.  

The plan of this paper is as follows: In Section \ref{Sec3}, we briefly introduce TEGR and its extensions like  $f(T)$ gravity and  $f(T,B)$ gravity.  $f(T,B)$  cosmology and particular cases that can be derived from it are  introduced in Section \ref{Sec4} . Section \ref{Sec5} is devoted to study of  Noether's symmetries for  $f(T,B)$ gravity. In particular, we derive the Noether vector field  and derive the Noether conditions for the function $f(T,B)$. In the related  subsections, we study particular cases of $f(T,B)$ function discussing, in particular, how it reduces to   $f(T)$, $f(R)$ and $f(R,T)$ gravities. The main point of this section is to demonstrate how several classes of modified gravity theories can be reduced to the $f(T,B)$ paradigm.  Discussion and conclusions are drawn in Section \ref{Sec6}. In our notation, Greek and Latin indices denote space-time and tangent coordinates respectively and  the signature $(+,-,-,-)$ is adopted for the metric.

\section{Teleparallel equivalent of general relativity and its modifications}

\label{Sec3}

We shortly present the basis of the teleparallel equivalent of general
relativity (TEGR) and its generalization, the so-called, $f(T)$ gravity. In
this theory, the vierbeins or tetrad fields $e_{\mu }^{a}$ are the dynamical
variables which form an orthonormal basis for the tangent space at each
point $x^{\mu }$ of the spacetime manifold. Hence, the tetrads $e_{\mu }^{n}$ and their
inverses $E_{m}^{\mu }$ obey the following orthogonality relations 
\begin{align}
E_{m}^{\mu }e_{\mu }^{n}& =\delta _{m}^{n}\,,  \label{deltanm} \\
E_{m}^{\nu }e_{\mu }^{m}& =\delta _{\mu }^{\nu }\,.  \label{deltamunu}
\end{align}%
Using the tetrad fields, the metric tensor can be constructed as
\begin{equation*}
g_{\mu \nu }=e_{\mu }^{a}e_{\nu }^{b}\eta _{ab}\,, 
\end{equation*}%
where $\eta _{ab}$ denotes the Minkowski metric. The main idea of TEGR is to construct a theory with a geometry
endorsed with torsion and having a globally flat curvature. To realize this program, we
define the torsion tensor by considering the curvatureless Weitzenb\"{o}ck
connection $W_{\mu }{}^{a}{}_{\nu }=\partial _{\mu }e_{\nu }^{a}$ such as 
\begin{equation}
T^{a}{}_{\mu \nu }=W_{\mu }{}^{a}{}_{\nu }-W_{\nu }{}^{a}{}_{\mu }=\partial
_{\mu }e_{\nu }^{a}-\partial _{\nu }e_{\mu }^{a}\,.
\end{equation}%
Additionally, it is convenient to define the contorsion tensor 
\begin{equation}
2K_{\mu }{}^{\lambda }{}_{\nu }=T^{\lambda }{}_{\mu \nu }-T_{\nu \mu
}{}^{\lambda }+T_{\mu }{}^{\lambda }{}_{\nu }\,,
\end{equation}%
and also the following tensor 
\begin{equation}
2S_{\sigma }{}^{\mu \nu }=K_{\sigma }{}^{\mu \nu }-\delta _{\sigma }^{\mu
}T^{\nu }+\delta _{\sigma }^{\nu }T^{\mu }\,.
\end{equation}%
The combination $S_{\sigma }{}^{\mu \nu }T^{\sigma }{}_{\mu \nu }$ is
denoted by $T$ and it is usually called the torsion scalar. This quantity is
a topological object and the TEGR is constructed by defining the
action 
\begin{equation}\label{TEGRac}
S_{\mathrm{TEGR}}=\frac{1}{\kappa }\int d^{4}x\,e\,T+S_{\mathrm{m}}\,,
\end{equation}%
where $S_{\mathrm{m}}$ denotes the action of any matter field and $e=\text{%
	det}(e_{\mu }^{a})=\sqrt{-g}$ is the volume element of the metric. The Ricci scalar $R$ and the torsion scalar $T$ differs by a
boundary term via, 
\begin{equation}
R=-T+\frac{2}{e}\partial _{\mu }(eT^{\mu })=-T+B\,.\label{RTB}
\end{equation}%
Here, for simplicity we  introduce $B=(2/e)\partial _{\mu }(eT^{\mu })=\nabla_{\mu}T^{\mu}$. 
We can easily see that due to the above relation, the TEGR action 
reproduces the same field equations as GR being \eqref{TEGRac} equivalent to the Hilbert-Einstein action.

Now, we can straightforwardly generalize  (\ref{TEGRac}) by
considering the following action 
\begin{equation}
S_{f(T)}=\frac{1}{\kappa }\int d^{4}x\,e\,f(T)+S_{\mathrm{m}}\,,
\label{action1}
\end{equation}%
where $f(T)$ is a smooth function of the torsion scalar. It is easy to see
that, by setting $f(T)=T$, the TEGR action is recovered. In this theory is
not possible to find the teleparallel equivalent of $f(R)$ gravity since now
the boundary term in (\ref{RTB}) contributes to the field equations.
Since $T$ itself is not invariant under local Lorentz transformations,
this theory is also not invariant under Lorentz transformations. An important fact is that this
theory is a second order one and hence, mathematically it is easier than $%
f(R)$ gravity.
The above action (\ref{action1}) can be immediately generalized by assuming that the function
$f(T)$ depends also on the boundary term $B$. The action reads as follows \cite%
{Bahamonde:2015zma} 
\begin{equation}\label{fTB}
S_{f(T,B)}=\frac{1}{\kappa }\int d^{4}x\,e\,f(T,B)+S_{\mathrm{m}}\,, 
\end{equation}%
where $f$ is a smooth function of two scalar fields, i.e. both the scalar torsion $T$ and the
boundary term $B$. The motivation of this
action comes out from the fact that from $f(T)$ gravity, it is not possible
to find an equivalent theory of its metric counterpart, the $f(R)$ gravity.
From the above action, we can easily see that the $f(R)$ and $f(T)$ can be
recovered by assuming  $f(T,B)=f(-T+B)=f(R)$ and $f(T,B)=f(T)$
respectively.

By varying the above action with respect to the tetrad field,  we get the
 field equations 
\begin{multline}
2eE_{a}^{\lambda }\Box f_{B}-2eE_{a}^{\sigma }\nabla ^{\lambda }\nabla
_{\sigma }f_{B}+eBE_{a}^{\lambda }f_{B}+4e\Big[(\partial _{\mu
}f_{B})+(\partial _{\mu }f_{T})\Big]S_{a}{}^{\mu \lambda } \\
+4\partial _{\mu }(eS_{a}{}^{\mu \lambda })f_{T}-4ef_{T}T^{\sigma }{}_{\mu
	a}S_{\sigma }{}^{\lambda \mu }-efE_{a}^{\lambda }=16\pi e\Theta
_{a}^{\lambda },  \label{FieldEQ}
\end{multline}%
where $f_{T}=\partial f/\partial T$, $f_{B}=\partial f/\partial B$, $\nabla
_{\sigma }$ is the covariant derivative with respect to the Levi-Civita
connection and $\Theta _{a}^{\lambda }$ is the energy-momentum tensor. 
As said before, this theory can summarize features of $f(T)$, $f(R)$ and $f(R,T)$ gravities.

\section{$f(T,B)$ cosmology}
\label{Sec4}
In this paper, we are interested in cosmological consequences of $f(T,B)$ gravity. In particular to find out exact cosmological solutions  by the Noether Symmetry Approach.
Let us consider  $f(T,B)$ cosmology in a flat Friedmann-Lema\^itre-Robertson-Walker (FLRW) universe. The
spatially flat FLRW metric in Cartesian coordinates reads as follows 
\begin{equation}
ds^{2}=dt^{2}-a(t)^{2}\Big(dx^{2}+dy^{2}+dz^{2}\Big)\,,  \label{FLRW}
\end{equation}%
where $a(t)$ is the scale factor of the universe. This metric can be
constructed by the following tetrad field 
\begin{equation}
e_{\mu }^{a}=\text{diag}\Big(1,a(t),a(t),a(t)\Big).  \label{tetrad1}
\end{equation}%
Since $f(T,B)$ is not invariant under Lorentz transformations, one needs to be very careful with
the choice of the tetrad. For instance, the unwanted condition $f_{TT}=0$
appears when one considers a flat diagonal FLRW tetrad in spherical
coordinates. The above vierbein is a \textquotedblleft good tetrad" in the
sense of Ref. \cite{Tamanini:2012hg} since it will not constraint our system.\newline
By considering a standard perfect fluid as a content of the universe and
using the above tetrad,   we find that the modified Friedmann equations are
given by 
\begin{eqnarray}  \label{fTB1}
-3H^{2}(3f_{B}+2f_{T})+3H\dot{f}_{B}-3\dot{H}f_{B}+\frac{1}{2}f(T,B)
&=&\kappa \rho (t)\,,  \label{equation1} \\
-(3H^{2}+\dot{H})(3f_{B}+2f_{T})-2H\dot{f}_{T}+\ddot{f}_{B}+\frac{1}{2}%
f(T,B) &=&-\kappa p(t)\,.  \label{equation2}
\end{eqnarray}%
Here dots represent derivation with respect to the cosmic time and $H=\dot{a}%
/a$ is the Hubble parameter. In addition, $\rho (t)$ and $p(t)$ are the
energy density and pressure of the cosmological fluid respectively. It is clear that by setting $f(T,B)=-f(T-B)=-f(-R)$ we  recover the FLRW equations in $f(R)$ gravity with the standard notation (see for example  \cite{rep2, Bahamonde:2016wmz}). Immediately  we have
\begin{eqnarray}
\label{JGRG15}
-\frac{f(R)}{2} + 3\left(H^2 + \dot H\right) f_{R}(R)
- 18 \left( 4H^2 \dot H + H \ddot H\right) f_{RR}(R)&=&\kappa \rho(t)\,,\\
\label{Cr4b}
\frac{f(R)}{2} - \left(\dot H + 3H^2\right)f_{R}(R)
+ 6 \left( 8H^2 \dot H + 4 {\dot H}^2 + 6 H \ddot H + \dddot H\right) f_{RR}(R)
\nonumber&&\\
+ 36\left( 4H\dot H + \ddot H\right)^2 f_{RRR}(R)&=&\kappa p(t)\,,
\end{eqnarray}
where $f_{R}=df(R)/dR$. Moreover, we can choose  $f(T,B)=f(T)$ to find the FLRW equations in $f(T)$ gravity given by
\begin{align}
12H^2f_{T}+f(T)&=2\kappa \rho(t)\,,\\
48H^2\dot{H}f_{TT}-(12H^2+4\dot{H})f_{T}-f(T)&=2\kappa p(t)\,,
\end{align}
see also \cite{Basilakos:2013rua}.
Note that the theory $f(R,T)$ can be viewed as a special case of $f(T,B)$ gravity since we can chose $f(T,B)=f(-T+B,T)=f(R,T)$. In this sense, it can be argued that $f(T,B)$ should be a more natural theory to consider than $f(R,T)$ as we will discuss below. 

In the specific case we are dealing with, the  cosmological equations can be derived both from the field Eqs.~(\ref{FieldEQ})  or deduced by a  point-like canonical Lagrangian ${\cal L}(a,{\dot a}, T,{\dot T},B,{\dot B})$ related to the action (\ref{fTB}), where dots represent derivation with respect to the cosmic time $t$.  Here,  ${\mathbb Q}\equiv\{a,T,B\}$ is the configuration space from which it is possible to derive 
${\mathbb{TQ}}\equiv \{a,\dot{a}, T, \dot{T}, B,{\dot B}\}$,  the corresponding tangent
space  on which ${\cal L}$ is defined as an application.  The variables $a(t)$,
$T(t)$ and $B(t)$ are, respectively,  the scale factor, the torsion scalar and the boundary term defined in the FLRW metric. 
The Euler-Lagrange equations are given by
\begin{eqnarray}
\frac{d}{dt}\frac{\partial {\cal L}}{\partial {\dot a}}=\frac{\partial {\cal L}}{\partial  a}\,, \qquad
\frac{d}{dt}\frac{\partial {\cal L}}{\partial {\dot T}}=\frac{\partial {\cal L}}{\partial  T}\,,\qquad
\frac{d}{dt}\frac{\partial {\cal L}}{\partial {\dot B}}=\frac{\partial {\cal L}}{\partial  B}\,,
\label{moto3}
\end{eqnarray}
with the energy condition
\begin{eqnarray}
E_{\cal L}= \frac{\partial {\cal L}}{\partial {\dot a}}{\dot a}+\frac{\partial {\cal L}}{\partial {\dot T}} {\dot T}+\frac{\partial {\cal L}}{\partial {\dot B}} {\dot B}-{\cal L}=0\,.
\label{energy}
\end{eqnarray}
As a consequence, the infinite number of degrees of freedom of the original field theory are
reduced to a finite number as in mechanical systems. 

Let us consider the canonical variables ${a,T,B}$ in order to derive the $f(T,B)$ action as follows
\begin{equation*}
S_{f(T,B)}=\int \mathcal{L}(a,{\dot{a}},T,\dot{T},B,\dot{B})dt \;.
\end{equation*}%
In a flat FLRW metric, it is 
\begin{align}
T&=-6\left[\frac{\dot{a}(t)}{a(t)}\right]^2\label{T}\, ,\\
B&=-6 \left[\frac{\ddot{a}(t)}{a(t)}+2\Big(\frac{ \dot{a}(t)}{a(t)}\Big)^2\right]\label{B}\,.
\end{align}
Therefore, the Ricci scalar is 
\begin{align}
R=-T+B=-6\left[\Big(\frac{\dot{a}(t)}{a(t)}\Big)^2+\frac{\ddot{a}(t)}{a(t)}\right]\,.
\end{align} 
By using (\ref{T}) and (\ref{B}), we can rewrite the action (\ref{fTB}) into its point-like  representation using the Lagrange multipliers $\lambda_{1}$ and $\lambda_{2}$ as
\begin{equation}
S_{f(T,B)}=2\pi ^{2}\int dt\left\{ (f(T,B))a^{3}-\lambda_{1} \left[ T+6\Big(\frac{\dot{a}}{a}\Big)^2\right] -\lambda_{2} \left( B+6 \left[\frac{\ddot{a}}{a}+2\Big(\frac{ \dot{a}}{a}\Big)^2\right]\right)\right\} \,.  \label{actioncan}
\end{equation}
By varying this action with respect to $T$ and $B$,  we find
\begin{align}
(a^3f_{T}-\lambda_{1}) \delta T&=0\ \rightarrow \lambda_{1}=a^3f_{T}\,,\\
(a^3f_{B}-\lambda_{2}) \delta B&=0 \ \rightarrow \lambda_{2}=a^3f_{B} \,.
\end{align}
Thus, the action (\ref{actioncan}) becomes
\begin{equation}
S_{f(T,B)}=2\pi ^{2}\int dt\left\{ (f(T,B))a^{3}-a^3f_{T} \left( T+6\Big(\frac{\dot{a}}{a}\Big)^2\right) -a^3f_{B} \left( B+6 \left[\frac{\ddot{a}}{a}+2\Big(\frac{ \dot{a}}{a}\Big)^2\right]\right)\right\} \,,  \label{actioncan2}
\end{equation}
and  the point-like Lagrangian is 
\begin{equation}
\mathcal{L}_{f(T,B)}=a^{3}\Big[ f(T,B)-Tf_T-Bf_{B}\Big]-6a\dot{a}^2f_{T}+6a^2\dot{a}\Big(f_{BT}\dot{T}+f_{BB}\dot{B}\Big) \,,
\label{L}
\end{equation}
where we have integrated by parts. This Lagrangian is  canonical and  depends on the three time-dependent  fields $a$, $T$,  and $B$. 
If we choose $f(T,B)=f(T)$,  we recover the teleparallel $f(T)$ cosmology with the Lagrangian  \cite{Basilakos:2013rua}
\begin{equation}
\mathcal{L}_{f(T)}=a^{3}\Big[ f(T)-Tf_T\Big]-6a\dot{a}^2f_{T}\,.
\label{Ltwo}
\end{equation}
In addition, if we choose $f(T,B)=f(-T+B)=f(R)$  we obtain the point-like Lagrangian action of $f(R)$ gravity  \cite{Capozziello:2008ch}
\begin{equation}
\mathcal{L}_{f(R)}=a^{3}\Big[ f(R)-Rf_R\Big]+6a\dot{a}^2f_{R}+6a^2\dot{a}\dot{R}f_{RR} \,.
\label{Ltthree}
\end{equation}
Moreover, we can  recover the teleparallel-curvature gravity assuming $f(T,B)=f(-T+B,T)=f(R,T)$  and   starting from the following considerations. In  this case,  we need to be careful in adopting the suitable variables. Assuming $x_{1}=-T+B=R$ and $x_{2}=T$, we have 
\begin{align}
 f_{T}&=\frac{\partial f}{\partial x_{1}}\frac{\partial x_{1}}{\partial T}+\frac{\partial f}{\partial x_{2}}\frac{\partial x_{2}}{\partial T}=-f_{x_{1}}+f_{x_{2}}=-f_{R}+f_{T}\,,\label{fT1}\\
 f_{B}&=\frac{\partial f}{\partial x_{1}}\frac{\partial x_{1}}{\partial B}+\frac{\partial f}{\partial x_{2}}\frac{\partial x_{2}}{\partial B}=f_{x_{1}}=f_{R}\,.\label{fT2}
\end{align}
Using the derivative chain rule, the second and third derivatives of $T$ and $B$ are  given by
\begin{align}
f_{TT}&=f_{RR}+f_{TT}-2f_{RT}\,,\label{fT3}\\
f_{TB}&=-f_{RR}+f_{RT}\,\label{fT4}\\
f_{BB}&=f_{RR}\,.\label{fT5}
\end{align}
The $f(R,T)$ point-like Lagrangian is given by
\begin{equation}
\mathcal{L}_{f(R,T)}=a^{3}\Big[ f(R,T)-Tf_T-Rf_{R}\Big]-6a\dot{a}^2(f_{T}-f_{R})+6a^2\dot{a}\Big(f_{RT}\dot{T}+f_{RR}\dot{R}\Big) \,.
\label{Lf(R,T)}
\end{equation}
see also \cite{Capozziello:2014bna} for a discussion.  With these considerations in mind, let us search for cosmological solutions for the above models by the Noether Symmetry Approach.

\section{Noether symmetry approach for $f(T,B)$ cosmology}\label{Sec5}

The Noether Symmetry Approach has been widely used  in the literature to find cosmological solutions in modified gravity (see \cite{cimento} for a comprehensive review). 
The main idea is to find symmetries in a given model  and then to use them to reduce related dynamical systems and find exact solutions. As a byproduct, the existence of the symmetries selects the functions inside the models (e.g. couplings and self-interaction potentials) that, in most cases, have a physical meaning. In this sense, the existence of a Noether symmetry is a sort of selection rule. Essentially, the technique consists in  deriving  constants of motions. 
Any constant of motion is related to a conserved quantity that allows to reduce the dynamical system and then to obtain exact solutions. If the number of constants is equal to the number of degrees of freedom, the system is completely integrable. 

In general, a  Noether symmetry for a given  Lagrangian  exists,  if the condition 
\begin{equation}
L_{X}\mathcal{L}=0 \;\; \rightarrow \;\; X \mathcal{L}=0,
\label{Lie}
\end{equation}
is satisfied.  $X$ is  the Noether vector field and $L_{X}$ is the Lie derivative. For  generalized coordinates $q_{i}$, we can construct the Noether vector field $X$. We have
\begin{equation}
X=\alpha^{i}(q)\frac{\partial }{\partial q^{i}}+\frac{d\alpha^{i}(q)}{dt}\frac{\partial}{\partial \dot{q}^{i}}\,,
\end{equation}
where $\alpha^i$ are functions defined in a given  configuration space  ${\mathbb Q}$ that assign the Noether vector. 
In our case, a symmetry generator $X$ in the space ${\mathbb Q}\equiv\{a,T,B\}$ is 
\begin{align}
X&=\alpha\partial_{a}+\beta \partial_{T}+\gamma\partial_{B}+\dot{\alpha}\partial_{\dot{a}}+\dot{\beta} \partial_{\dot{T}}+\dot{\gamma}\partial_{\dot{B}},
\end{align}
where $\alpha,\beta,\gamma$ depend on $a$, $T$ and $B$. Therefore we have
\begin{align}
\dot{\alpha}&=\Big(\frac{\partial \alpha}{\partial a}\Big)\dot{a}+\Big(\frac{\partial \alpha}{\partial T}\Big)\dot{T}+\Big(\frac{\partial \alpha}{\partial B}\Big)\dot{B}\,,\\
\dot{\beta}&=\Big(\frac{\partial \beta}{\partial a}\Big)\dot{a}+\Big(\frac{\partial \beta}{\partial T}\Big)\dot{T}+\Big(\frac{\partial \beta}{\partial B}\Big)\dot{B}\,,\\
\dot{\gamma}&=\Big(\frac{\partial \gamma}{\partial a}\Big)\dot{a}+\Big(\frac{\partial \gamma}{\partial T}\Big)\dot{T}+\Big(\frac{\partial \gamma}{\partial B}\Big)\dot{B}\,.
\end{align}
A Noether symmetry exists if at least one of the functions $\alpha$, $\beta$, and $\gamma$ is different from zero. Their analytic forms can be found by making explicit Eq. (\ref{Lie}), which corresponds to a set of partial differential equations given by equating to zero the terms in $\dot{a}^{2},\dot{a}\dot{T},\dot{a}\dot{B},\dot{T}^{2},\dot{B}^{2},\dot{B}\dot{T}$ and so on. For a $n$ dimensional configuration space,  we  have $1+n(n+1)/2$ equations derived from Eq. (\ref{Lie}).  In our case, the configuration space  is three dimensional, so we  have seven partial differential equations. Explicitly, from  (\ref{Lie}),  we find the following system of partial differential equations
\begin{align}
f_{T} \left(2 a \frac{\partial \alpha}{\partial a}+\alpha\right)+f_{TB} \left(a \gamma-a^2 \frac{\partial \beta}{\partial a}\right)+a f_{TT} \beta-a^2 f_{BB} \frac{\partial \gamma}{\partial a}&=0\,,\label{1}\\
f_{TB} \left(a \frac{\partial \alpha}{\partial a}+a\frac{\partial \beta}{\partial T}+2 \alpha\right)-2  f_{T}\frac{\partial \alpha}{\partial T}+a  f_{BB} \frac{\partial \gamma}{\partial T}+a  (f_{TTB} \beta+f_{TBB} 
\gamma)&=0\,,\label{2}\\
f_{BB} \left(a \frac{\partial \alpha}{\partial a}+a \frac{\partial \gamma}{\partial B}+2 \alpha\right)+a f_{TB} \frac{\partial \beta}{\partial B}+a (\beta f_{TBB} + \gamma f_{BBB} )-2 f_{T} \frac{\partial \alpha}{\partial B}&=0\,,\label{3}\\
f_{TB}\frac{\partial \alpha}{\partial T}&=0\,,\label{4}\\
f_{BB}\frac{\partial \alpha}{\partial B}&=0\,,\label{5}\\
f_{TB}\frac{\partial \alpha}{\partial B}+f_{BB}\frac{\partial \alpha}{\partial T}&=0\,\label{6}\\
3 \left(f-B f_{B}-T f_{T}\right) \alpha -a \left(B f_{TB}+Tf_{TT}\right) \beta -a \left(B f_{BB}+T f_{TB}\right) \gamma &=0\,.\label{7}
\end{align}
where the unknown variables are $\alpha$, $\beta$, $\gamma$ and the function  $f(T,B)$.
 There are two different strategies to solve it and  to find  symmetries: $(i)$  one can directly  solve the system  (\ref{1})-(\ref{7}) and then find the unknown functions; $(ii)$ one can impose specific forms of  $f(T,B)$ and search for the related  symmetries \cite{kostas}.
 From a physical viewpoint, the second approach is better because it allows to  study reliable models. By the first strategy, solutions can be achieved but, in most cases, they are implicit functions that do not allow a physical analysis \cite{cimento}.
 We will adopt the second one to discuss the $f(T,B)$ cosmology.
  
\subsection{Case 1: $f(T,B)=b_{0}B^{k}+t_{0}T^{m}$}
 For a power-law like function given by $f(T,B)=b_{0}B^{k}+t_{0}T^{m}$, where $b_{0}$, $t_{0}$, $k$ and $m$ are constants, we  find that the unique solution of (\ref{1})-(\ref{7}) is for  $k=1$. This is  trivial because it gives $f(T,B)=b_{0}B+t_{0}T^{m}$ which is the same as a power-law $f(T)$ function. This  comes from the fact that $B$ is a boundary term so that a linear form of the function in $B$ does not introduce any change in the field equations. Hence, this kind of function gives the same results reported in  \cite{Basilakos:2013rua}.
 
\subsection{Case 2: $f(T,B)=f_{0}B^{k}T^{m}$}
Let us now study the case where the function takes the  form
\begin{eqnarray}
\label{powerB}
f(T,B)&=&f_{0}B^{k}T^{m}\,,
\end{eqnarray}
where $f_{0}$, $k$ and $m$ are constants. 
From (\ref{4})-(\ref{6}), it is  $\alpha=\alpha(a)$.
If we replace the function \eqref{powerB} into (\ref{1})-(\ref{7}),  we find the following Noether's vector:
\begin{eqnarray}
X=\frac{\alpha_{0}}{a^2}\partial_{a}-\frac{6\alpha_{0} T}{a^3}\partial_{T}-\frac{3\alpha_{0}B}{a^3}\partial_{B}\,,
\end{eqnarray}
and also the constraint $k=1-m$ which gives us $f(T,B)=f_{0}B^{k}T^{\frac{1-k}{2}}$. Here, $\alpha_{0}$ is an integration constant that can be set equal to 1 without loss of generality \cite{cimento}. Let us now find out cosmological solutions for this type of function. The point-like Lagrangian (\ref{L}) takes the following form
\begin{equation}
\mathcal{L}=\frac{1}{2} f_{0}(k-1) a(t) B^{k-2} T^{-\frac{1}{2} (k+1)} \left[6 B^2 \dot{a}(t)^2-6 k a(t) \dot{a}(t) (B\dot{T} -2 \dot{B} T)-B^2 T a(t)^2\right]\,.
\label{L22}
\end{equation}
It is easy to see that the trivial case $k=1$, which produces $f=f_{0}B$, gives the expected result where the field equations are identically zero. The Euler-Lagrange equation for the scale factor $a(t)$ gives, for $k\neq1$ and $f_{0}\neq 0$, 
\begin{eqnarray}
a(t)^2 \Big(4 (k-2) k T^2 \dot{B}^2+4 k B T \left(T \ddot{B}-(k-1) \dot{B}\dot{T}\right)+k B^2 \left((k+1) \dot{T}^2-2 T \ddot{T}\right)+B^3 T^2\Big)\nonumber\\
+2 B^3 T \dot{a}(t)^2+2 a(t) B^2 \left(2 B T \ddot{a}(t)+\dot{a}(t) \left(2 k T \dot{B}-(k+1) B \dot{T}\right)\right)=0\,.\label{lls2}
\end{eqnarray}
Additionally, the energy equation becomes
\begin{eqnarray}
-6 B^2 \dot{a}(t)^2+6 k a(t) \dot{a}(t) (B\dot{T} -2 \dot{B} T)+B^2 T a(t)^2=0\,.\label{lls}
\end{eqnarray}
If we replace $T$ and $B$ given by (\ref{T}) and (\ref{B}) we find that Eqs. (\ref{lls2}) and (\ref{lls}) become
\begin{eqnarray}
(k-1) a(t)^4 \ddot{a}(t)^4+4 (k-2) \dot{a}(t)^8-4 (k-4) a(t)^2 \dddot{a}(t) \dot{a}(t)^5-8 (k-1) a(t) \dot{a}(t)^6 \ddot{a}(t)\nonumber\\
+4 (k-2) a(t)^3 \dddot{a}(t) \dot{a}(t)^3 \ddot{a}(t)+2 a(t)^2 \dot{a}(t)^4 \left(a(t) \ddddot{a}(t)+2 (2 k-5) \ddot{a}(t)^2\right)\nonumber\\
+a(t)^3 \dot{a}(t)^2 \left((k-2) a(t) \dddot{a}(t)^2+(10-4 k) \ddot{a}(t)^3+a(t) \ddddot{a}(t) \ddot{a}(t)\right)\nonumber\\
-2 (k-1) a(t)^4 \dddot{a}(t) \dot{a}(t) \ddot{a}(t)^2=0\,,\\
-(k-1) a(t)^2 \ddot{a}(t)^2-2 (k-2) \dot{a}(t)^4+k a(t)^2 \dddot{a}(t) \dot{a}(t)+2 (k+2) a(t) \dot{a}(t)^2 \ddot{a}(t)=0\,.
\end{eqnarray}
These equations admit power law solutions for the scale factor being
\begin{eqnarray}
a(t)=a_{0}t^{s}\,, \ \ \ s=\frac{1+k}{3}\,.
\end{eqnarray}
The torsion scalar  and the boundary term are $T=-6s^2/t^2$ and $B=6s(1-3s)/t^2$ respectively.  Immediately we see that several cosmologically interesting cases can be recovered.
 A  radiation  solution  is for 
 \begin{equation}
 a(t) = a_0 t^{1/2},\;\;\;\; \mbox{with}\;\;\;\;k=\frac{1}{2}\,.
   \end{equation}
   A dust solution is for 
\begin{equation}
 a(t) = a_0 t^{2/3},\;\;\;\; \mbox{with}\;\;\;\;k=1\,.
   \end{equation} 
A stiff matter one is for
 \begin{equation}
 a(t) = a_0 t^{1/3},\;\;\;\; \mbox{with}\;\;\;\;k=0\,.
   \end{equation} 
Power-law inflation is recovered for $s\geq1$ and  $k\geq 2$.   

\subsection{Case 3: $f(T,B)=-T+F(B)$}\label{fff}
The  case  $f(T,B)=-T+F(B)$ is a deviation of TEGR up to a function which depends on the boundary term.
The Noether condition gives
\begin{align}
	2 a \frac{\partial \alpha}{\partial a}+\alpha+a^2 F_{BB} \frac{\partial \gamma}{\partial a}&=0\,,\label{1f}\\
	2  \frac{\partial \alpha}{\partial T}+ a  F_{BB} \frac{\partial \gamma}{\partial T}&=0\,,\label{2f}\\
	F_{BB} \left(a \frac{\partial \alpha}{\partial a}+a \frac{\partial \gamma}{\partial B}+2 \alpha\right)+a \gamma F_{BBB}+2  \frac{\partial \alpha}{\partial B}&=0\,,\label{3f}\\
	F_{BB}\frac{\partial \alpha}{\partial B}&=0\,,\label{5f}\\
	F_{BB}\frac{\partial \alpha}{\partial T}&=0\,,\label{6f}\\
	3 \alpha \left(f(B)-B F_{B}\right) -a B F_{BB} \gamma &=0\,.\label{7f}
\end{align}
 Discarding the trivial case  $F(B)=B$ which gives standard TEGR, from (\ref{5f}) and (\ref{6f}) we obtain again that  
	$\alpha=\alpha(a)$.
Using this condition in (\ref{2f}), we  find that $\gamma=\gamma(B,a)$ and the  equations become
\begin{align}
	2a \frac{d \alpha}{d a} + \alpha-a^2 F_{BB}\frac{\partial \gamma}{\partial a}&=0,\label{1g}\\
	F_{BB}\Big(a \frac{d \alpha}{da}+a \frac{\partial \gamma}{\partial B}+2\alpha\Big)+a\gamma F_{BBB}&=0,\label{3g}\\
	3 \alpha \left(F(B)-B F_{B}\right) -a B F_{BB} \gamma &=0\,.\label{7g}
\end{align}
We can rewrite (\ref{3g}) as
\begin{align}
	\partial_{B}(\gamma F_{BB})&=- F_{BB}\Big(\frac{d \alpha}{da}+2\frac{\alpha}{a}\Big)\,,
\end{align}
which can solved for $\gamma$, yielding 
\begin{align}
	\gamma&=-\Big(\frac{d \alpha}{da}+2\frac{\alpha}{a}\Big)\frac{F_{B}}{F_{BB}}+\frac{g(a)}{F_{BB}}\,,\label{gammapr}
\end{align}
where $g(a)$ is an arbitrary function of the scale factor. Therefore, from (\ref{1g}) one find that
\begin{align}
	F_{B}\Big(2\alpha-2a \frac{d \alpha}{da}-a^2 \frac{d^2 \alpha}{da^2}\Big)+\alpha+2a\frac{d \alpha}{da}+a^2\frac{d g}{da} &=0\,,
\end{align}
which has the following solution
\begin{align}
	\alpha(a)&=c_{1}a+\frac{C_{2}}{a^2} \,, \, \, \, g(a)=c_{3}-\frac{C_{2}}{a^3}-3c_{1}\log{a}\,.
\end{align}
Here, $c_{1}$, $C_{2}$ and $c_{3}$ are integration constants. Now, by using (\ref{7g}) and (\ref{gammapr}) we finds that
\begin{align}
	a^3 (3 B c_{1} \log (a)-B c_{3}+3 c_{1} F)+C_{2} (B+3 F-3 B F_{B})=0\,.
\end{align}
Since $F=F(B)$, we  have that the first term is zero, so that $c_{1}=c_{3}=0$, yielding
\begin{align}
B+3 F-3 B F_{B}=0\,,
\end{align}
which can be solved obtaining
\begin{align}
	F(B)&=f_{0}B +\frac{1}{3} B \log (B)\,.
\end{align}
Therefore, we find the following symmetry solutions
\begin{eqnarray}
X&=&\frac{C_{2}}{a^2}\partial_{a}+\beta\partial_{T}-\frac{C_{2}}{a^3F_{BB}}\partial_{B}\,,\label{XX}\\
f(T,B)&=&T+f_{0}B +\frac{1}{3} B \log (B)\,.
\end{eqnarray}
Let us now search for   cosmological solutions for this model. Considering (\ref{XX}), it is convenient to introduce the following  coordinates 
\begin{equation}
u=\frac{1}{3C_{2}}a^3\,,~v=\frac{1}{3C_{2}}\Big[F_{B}+\log (a)\Big]\,,\label{uv}
\end{equation}%
which transform the Noether vector as 
 \begin{align} X=\partial_{u}+\beta \partial_{T}\,. 
 \end{align} 
 Lagrangian (\ref{L}) reads as follows 
\begin{eqnarray}
\mathcal{L}=\frac{2C_{2} }{\ddot{u}(t)}\left[\ddot{u}(t)^2+\dddot{u}(t) \dot{u}(t)\right]\,,
\end{eqnarray}
and hence, the Euler-Lagrange equation for $u(t)$ is
\begin{eqnarray}
\ddddot{u}(t)-\frac{\dddot{u}(t)^2}{\ddot{u}(t)}&=&0\,.
\end{eqnarray}
Hence, it is easily to find the following solution
\begin{eqnarray}
u(t)&=&\frac{u_{3} }{u_{1}^2}e^{u_{1}t}+u_{2}t +u_{0}\,,
\end{eqnarray}%
where $u_{0},u_{1},u_{2}$ and $u_{3}$ are integration constants.
Additionally, since $\mathcal{L}=E-2V$, with $E$ being the Hamiltonian (the energy) of the system and $V(t)=2C_{2} u_{3} e^{t u_{1}}$ can be understood as an energy potential, we find the following constraint  
\begin{eqnarray}
2C_{2}u_{1}v_{1}&=&E\,.
\end{eqnarray}
Finally, using (\ref{uv}) we can express this cosmological solution in term of the scale factor as follows,
\begin{eqnarray}
a(t)&=&\Big[\frac{3 C_{2}u_{3} e^{u_{1}t}}{u_{1}^2}+3 C_{2}\Big( t u_{2}+u_{0}\Big)\Big]^{1/3}\,.
\end{eqnarray}
It is easy to see that this solution  gives a de Sitter universe for the specific choice $u_{2}=u_{0}=0$. This de Sitter solution is reported also in \cite{Bahamonde:2016cul} where a  cosmological reconstruction technique is adopted.
	
In next subsections,  remarkable theories that can be recovered from $f(T,B)$ gravity are discussed. We will see that all  symmetries found in earlier studies for $f(T)$, $f(R)$ and $f(R,T)$ can be achieved starting from  the Noether symmetry equations Eqs.(\ref{1})-(\ref{7}) derived for   $f(T,B)$ cosmology.

\subsection{Case 4: $f(T,B)=f(T)$}
The first remarkable example is $f(T)$ gravity. 
The cases  studied in \cite{Atazadeh:2011aa,Wei:2011aa} are straightforwardly obtained. Eqs. (\ref{4})-(\ref{6}) are identically satisfied since $f_{TB}=f_{BB}=0$. The other equations become
\begin{align}
f_{T} \left(2 a \frac{\partial \alpha}{\partial a}+\alpha\right)+a f_{TT} \beta&=0\,,\label{1b}\\
 f_{T}\frac{\partial \alpha}{\partial T}&=0\,,\label{2b}\\
f_{T} \frac{\partial \alpha}{\partial B}&=0\,,\label{3b}\\
3 \left(f-T f_{T}\right) \alpha -a\beta T f_{TT} &=0\,.\label{7b}
\end{align}
By discarding the TEGR case ($f(T)=-T$) we have that $f_{T}\neq0$ and hence, from Eqs. (\ref{2b}) and (\ref{3b}),  we find 
again $\alpha=\alpha(a)$.
From Eq. (\ref{7b}),  we find that
\begin{align}
\alpha(a)&=\frac{af_{TT}T}{3(f-T f_{T})}\beta(a,T,B)\,
\end{align}
By replacing this expression in (\ref{1b}) we get the following differential equation for $\beta$
\begin{align}
\frac{\partial \beta}{\partial a}&=-\frac{3 f }{2 a f_{T} T}\beta(a,T,B)\,. 
\end{align}
To solve this equation, let us assume  that $\beta$ can be separated as $\beta(a,T,B)=\beta_{1}(a)\beta_{2}(T)\beta_{3}(B)$. We obtain
\begin{align}
\frac{2a}{\beta_{1}}\frac{d \beta_{1}}{d a}&=-\frac{3 f }{ f_{T} T}=-\frac{3}{C}\,. 
\end{align}
Here we have used that the l.h.s of the equation only depends on $a$ and the r.h.s only on $T$, so that $C$ is a constant. Thus, it is easy to solve the above equation yielding
\begin{eqnarray}
f(T)&=&f_{0}T^{C}\,,
\end{eqnarray}
where $f_{0}$ is an integration constant. Moreover, it is straightforward to find that the Noether symmetry vector becomes
\begin{eqnarray}
X&=&-\frac{1}{3} \beta_{0} a^{1-\frac{3}{2 C}}\partial_{a}+\frac{\beta_{0} T a^{-\frac{3}{2 C}}}{C}\partial_{T}+\gamma \partial_{B}\,,
\end{eqnarray}
where $\beta_{0}$ is an integration constant. As shown  in \cite{Atazadeh:2011aa,Wei:2011aa},  using this symmetry, one  finds that $f(T)$ gravity admits power-law cosmological solutions of the form of $a(t)\propto t^{-2C/C_{3}}$. A more general study of power-law $f(T)$ cosmology is in \cite{Basilakos:2013rua}.

\subsection{Case 5: $f(T,B)=f(-T+B)=f(R)$}
We can recover $f(R)$ gravity by assuming $f(T,B)=f(-T+B)=f(R)$. Hence, $f_{R}(R)=f'(-T+B)=-f_{T}=f_{B}$ and the  system of differential equations (\ref{1})-(\ref{7}) related to the Noether's symmetry in $f(R)$ gravity becomes
\begin{align}
f_{R} \left(2 a \frac{\partial \alpha}{\partial a}+\alpha\right)-af_{RR} \left(\beta+a \frac{\partial \beta}{\partial a} -\gamma-a\frac{\partial \gamma}{\partial a}\right) &=0\,,\label{1c}\\
-f_{RR} \left(a \frac{\partial \alpha}{\partial a}+a\frac{\partial \beta}{\partial T}+2 \alpha-a   \frac{\partial \gamma}{\partial T}\right)+2  f_{R}\frac{\partial \alpha}{\partial T}+a f_{RRR} ( \beta- \gamma)&=0\,,\label{2c}\\
f_{RR} \left(a \frac{\partial \alpha}{\partial a}+a \frac{\partial \gamma}{\partial B}+2 \alpha-a  \frac{\partial \beta}{\partial B}\right)+a  f_{RRR}(\gamma-\beta  )+2 f_{R} \frac{\partial \alpha}{\partial B}&=0\,,\label{3c}\\
-f_{RR}\frac{\partial \alpha}{\partial T}&=0\,,\label{4c}\\
f_{RR}\frac{\partial \alpha}{\partial B}&=0\,,\label{5c}\\
-f_{RR}\Big(\frac{\partial \alpha}{\partial B}-\frac{\partial \alpha}{\partial T}\Big)&=0\,\label{6c}\\
3 \alpha \left(f-Rf_{R}\right) +a Rf_{RR}( \beta - \gamma) &=0\,.\label{7c}
\end{align}
In addition, we require that $\beta=-\gamma$ to obtain the same generators as in $f(R)$ gravity. In doing this, Eqs. (\ref{2c}) and (\ref{3c}) are identical and hence the Noether equations become
\begin{align}
f_{R} \left(2 a \frac{\partial \alpha}{\partial a}+\alpha\right)+2af_{RR} \left(\gamma+a\frac{\partial \gamma}{\partial a}\right) &=0\,,\label{11}\\
f_{RR}\frac{\partial \alpha}{\partial B}&=0\,.\\
f_{RR}\Big(a \frac{\partial \alpha}{\partial a}+2\alpha+2a\frac{\partial \gamma}{\partial R}\Big)+2f_{R}\frac{\partial \alpha}{\partial R}+2a\gamma f_{RRR}&=0\,\\
3 \alpha \left(f-Rf_{R}\right) -2a\gamma Rf_{RR} &=0\,.\label{77}
\end{align}
It is worth noticing that, in order to recover the same Noether symmetry equations as in \cite{Capozziello:2008ch}, we require that $\gamma=\frac{1}{2}\tilde{\beta}$. This issue comes out in the computation of the Lie derivative since the generator and some terms related with the generator of $T$ and $B$ are summed twice. Therefore, by changing $\gamma=\frac{1}{2}\tilde{\beta}$ we find the same equations as in \cite{Capozziello:2008ch}, that is 
\begin{align}
f_{R} \left(2 a \frac{\partial \alpha}{\partial a}+\alpha\right)+af_{RR} \left(\tilde{\beta}+a\frac{\partial \tilde{\beta}}{\partial a}\right) &=0\,,\label{eq1again}\\
f_{RR}\Big(a \frac{\partial \alpha}{\partial a}+2\alpha+a\frac{\partial \tilde{\beta}}{\partial R}\Big)+2f_{R}\frac{\partial \alpha}{\partial R}+a\tilde{\beta} f_{RRR}&=0\,\label{eq2again}\\
f_{RR}\frac{\partial \alpha}{\partial R}&=0\,,\label{eq3again}\\
3 \alpha \left(f-Rf_{R}\right) -a Rf_{RR} \tilde{\beta} &=0\,.\label{eq4again}
\end{align}
Since we are not interested on the GR case, $f_{RR}\neq 0$ and, from (\ref{eq3again}),  we directly find that $\alpha=\alpha(a)$. Hence, Eq. (\ref{eq2again}) can be rewritten as 
\begin{eqnarray}
\partial_{R}(\beta f_{RR})&=&-f_{RR}\Big(\frac{d\alpha}{da}+\frac{2\alpha}{a}\Big)\,,
\end{eqnarray}
and solved yielding
\begin{eqnarray}
\beta(a,R)=\frac{g(a)}{f_{RR}(R)}-\frac{\left(a \alpha'(a)+2 \alpha (a)\right) f_{R}(R)}{a f_{RR}(R)}\,,
\end{eqnarray}
where $g(a)$ is an arbitrary function depending on $a$. Note that the latter solution is very similar to the one found  in (\ref{gammapr}) for the case $f(T,B)=-T+F(B)$. Now if we replace this solution into (\ref{eq1again}),  we obtain
\begin{align}
	f_{R}(R) \left[\alpha (a)-a \left(a \alpha''(a)+\alpha '(a)\right)\right]+a \left[a g'(a)+g(a)\right]=0\,,
	\end{align}
which is satisfied only if each bracket is zero. We have
\begin{eqnarray}
\alpha(a)&=&\frac{\left(a^2+1\right) \alpha_0}{2 a}-\frac{\left(a^2-1\right) \alpha_1}{2 a}\,,\\
g(a)&=&\frac{c}{a}\,,
\end{eqnarray}
where $c,\alpha_{0}$ and $\alpha_1$ are integration constants. It is important to mention that this result is more general than that in  \cite{Capozziello:2008ch} where some terms in  $\alpha(a)$ are not present;  however  the final result does not changes since the symmetry vectors are similar. By replacing the above expression into (\ref{eq4again}), we find
\begin{eqnarray}
\frac{(\alpha_0+\alpha_1) \left(3 f(R)-2 R f_R(R)\right)}{2 a}+\frac{3}{2} a (\alpha_{0}-\alpha_1) f(R)-c R=0\,,
\end{eqnarray}
which is valid only if $c=0$ and $\alpha_{0}=\alpha_{1}$. This   gives the result
\begin{eqnarray}
f(R)&=&f_{0}R^{3/2}\,,
\end{eqnarray}
where $f_{0}$ is an integration constant. By   considerations similar to those in  Sec. \ref{fff}, it is possible to show that $f(R)$ gravity admits power-law solution of the form 
\begin{align}
a(t)\propto t^{1/2}\,,\quad  \mbox{and} \quad a(t) = a_0[c_4t^4 + c_3t^3 + c_2t^2 + c_1t + c_0]^{1/2}\,.
\end{align}
For a discussion on the physical meaning of such solutions, see \cite{cap}.

\subsection{Case 6: $f(T,B)=f(-T+B,T)=f(R,T)$}
Let us now discuss the case where $f(T,B)=f(-T+B)=f(R,T)$. First of all, in order to have the same generator as in $f(R,T)$, we require to change the function $\gamma(T,B,a)\rightarrow \gamma(T,B,a)+\beta(T,B,a)$. Additionally, for the derivative terms,  we need to use Eqs.~(\ref{fT1})-(\ref{fT5}) and hence  the transformation $\partial/\partial T \rightarrow \partial/\partial T-\partial/\partial R$. After these replacements, the Noether conditions become
\begin{eqnarray}
\alpha(f_{R}-f_{T})+a\gamma(f_{RR}-f_{RT})+a\beta(f_{RT}-f_{TT})+2a\frac{\partial \alpha}{\partial a}(f_{R}-f_{T})+a^2f_{RR}\frac{\partial \gamma}{\partial a}\nonumber\\
+a^2f_{RT}\frac{\partial \beta}{\partial a}=0\,,\label{1d}\\
(f_{RT}-f_{RR}) \left(a \frac{\partial \alpha}{\partial a}+a\frac{\partial \beta}{\partial T}-a\frac{\partial \beta}{\partial R}+2 \alpha\right)-2  (f_{T}-f_{R})\Big(\frac{\partial \alpha}{\partial T}-\frac{\partial \alpha}{\partial R}\Big)\nonumber\\+a  f_{RR} \Big(\frac{\partial (\gamma+\beta)}{\partial T}-\frac{\partial (\gamma+\beta)}{\partial R}\Big)+a  ((f_{TRR}-f_{RRR}) (\beta+\gamma)\nonumber \\
+(f_{RRR}+f_{TTR}-2f_{TRR})\beta )=0\,,\label{2d}\\
f_{RR} \left(a \frac{\partial \alpha}{\partial a}+a \frac{\partial (\beta+\gamma)}{\partial R}+2 \alpha\right)+a (f_{RT}-f_{RR}) \frac{\partial \beta}{\partial R}+a (\beta(f_{TRR}-f_{RRR}) + (\gamma+\beta) f_{RRR} )\nonumber\\
-2 (f_{T}-f_{R}) \frac{\partial \alpha}{\partial R}=0\,,\label{3d}\\
(f_{RT}-f_{RR})\Big(\frac{\partial \alpha}{\partial T}-\frac{\partial \alpha}{\partial R}\Big)=0\,,\label{4d}\\
f_{RR}\frac{\partial \alpha}{\partial R}=0\,,\label{5d}\\
(f_{RT}-f_{RR})\frac{\partial \alpha}{\partial R}+f_{RR}\Big(\frac{\partial \alpha}{\partial T}-\frac{\partial \alpha}{\partial R}\Big)=0\,,\label{6d}\\
3\alpha \left(f-R f_{R}-T f_{T}\right)-a\gamma(Tf_{RT}+Rf_{RR})  -a \beta(Tf_{TT}+Rf_{RT}) =0\,.\label{7d}
\end{eqnarray}
By adding (\ref{2d}) with  (\ref{3d}),  we get
\begin{eqnarray}
2 \alpha  f_{RT}+a \gamma  f_{TRR}+a \beta  f_{TTR}+a f_{RT} \frac{\partial \alpha}{\partial a}+2f_{R} \frac{\partial \alpha}{\partial T} -2f_{T} \frac{\partial \alpha}{\partial T}+a f_{RR}\frac{\partial \gamma}{\partial T}+af_{RT}\frac{\partial \beta}{\partial T}  &=&0\,.
\end{eqnarray}
In addition, by subtracting (\ref{4d}) with (\ref{6d}) and using (\ref{5d}) and then adding  (\ref{4d}) with (\ref{6d}), we get $f_{RT}\frac{\partial \alpha}{\partial T}=0$ and $f_{RT}\frac{\partial \alpha}{\partial T}=0$. Therefore, the Noether symmetry equations can be rewritten as follows
\begin{eqnarray}
\alpha(f_{R}-f_{T})+a\gamma(f_{RR}-f_{RT})+a\beta(f_{RT}-f_{TT})+2a\frac{\partial \alpha}{\partial a}(f_{R}-f_{T})+a^2f_{RR}\frac{\partial \gamma}{\partial a}+a^2f_{RT}\frac{\partial \beta}{\partial a}&=&0\,,\nonumber\\\label{11d}\\
2 \alpha  f_{RT}+a \gamma  f_{TRR}+a \beta  f_{TTR}+a f_{RT} \frac{\partial \alpha}{\partial a}+2f_{R} \frac{\partial \alpha}{\partial T} -2f_{T} \frac{\partial \alpha}{\partial T}+a f_{RR}\frac{\partial \gamma}{\partial T}+af_{RT}\frac{\partial \beta}{\partial T}&=&0\,,\nonumber\\\label{22d}\\
2 \alpha f_{RR}+a f_{RRR}\gamma +a f_{TRR}\beta+a f_{RR}\frac{\partial \alpha}{\partial a} +2f_{R}\frac{\partial \alpha}{\partial R}-2f_{T}\frac{\partial \alpha}{\partial R} +a f_{RR}\frac{\partial \gamma}{\partial R}+af_{RT} \frac{\partial \beta}{\partial R} &=&0\,,\nonumber\\\label{33d}\\
f_{RT}\frac{\partial \alpha}{\partial T}&=&0\,,\nonumber\\\label{44d}\\
f_{RR}\frac{\partial \alpha}{\partial R}&=&0\,,\nonumber\\\label{55d}\\
f_{RT}\frac{\partial \alpha}{\partial R}&=&0\,,\nonumber\\ \label{66d}
3\alpha \left(f-R f_{R}-T f_{T}\right)-a\gamma(Tf_{RT}+Rf_{RR})  -a \beta(Tf_{TT}+Rf_{RT}) &=&0\,.\nonumber\\\label{77d}
\end{eqnarray}
It is clear that by changing $\beta\rightarrow \gamma$ and  $\gamma\rightarrow \beta$, the system of differential equations (\ref{11d})-(\ref{77d}) for the  Noether symmetry of $f(R,T)$ result the same as those  studied in \cite{Capozziello:2014bna} with the same physical implications. 

\section{Discussion}\label{Sec6}
In this paper, we discussed an extension of modified teleparallel gravity including  functions of the torsion scalar $T$ and its related boundary term $B=\frac{2}{e}\partial_{\mu}(e T^{\mu})$. In such a way, a gravitational theory with two fields, i.e. $T$ and $B$,  can be taken into account. If not assumed in a trivial way, that is linear in $B$, interesting features come out from the combinations of $T$ and $B$, in particular, the possibility to relate $f(T)$ and $f(R)$ gravity under the same standard. This means that not only GR and TEGR (respectively theories  linear in the Ricci scalar $R$ and the torsion scalar $T$ in their actions) result the "same" effective  theory but  also their extensions, also if conceptually very different, can show analogies and similitudes. 

Here we consider the Noether Symmetry Approach in order to investigate the related cosmologies. The main result is that the Noether vector fields emerging from $f(T,B)$ gravity are a general standard to find out solutions both in the starting theory and in the particular cases like $f(T)$, $f(R)$, and $f(R,T)$. In this last case, the Noether technique allows to deal with curvature $R$ and torsion $T$ scalars as two scalar fields. 

The related cosmological solutions are of physical interest and, essentially, all the main cosmological behaviors can be recovered.
However, this is only a preliminary study where no effective comparison with observations has been made and only toy models have been analyzed in order to test the technique. 

In  forthcoming  studies, we will adopt an  approach for $f(T,B)$ gravity as in \cite{Basilakos:2013rua}, where the condition (\ref{Lie}) is extended  to the possibility of discussing singular Lagrangians. Furthermore, a similar approach can be used for   teleparallel modified Gauss-Bonnet gravity $f(T,B,T_{G},B_{G})$ as studied in \cite{Bahamonde:2016kba}. Under this standard, other  interesting models  can  naturally arise by taking into account some specific functions of $T$ and $B$ as $f=f(-T+B,-T_{G}+B_{G})=f(R,G)$ (modified Gauss-Bonnet) or $f=f(T,T_{G})$ gravity (modified teleparallel Gauss-Bonnet). The final issue is to define a {\it mother theory} by which all  extensions and modifications of GR can be generated. 

\begin{flushleft}
    \textbf{Acknowledgments}
\end{flushleft}
S.B. is supported by the Comisi{\'o}n Nacional de Investigaci{\'o}n
Cient{\'{\i}}fica y Tecnol{\'o}gica (Becas Chile Grant
No.~72150066). S.C. acknowledges the financial support of INFN ({\it iniziative specifiche} TEONGRAV and QGSKY).
This article is based upon work from COST Action CA15117 (CANTATA), supported by COST (European Cooperation in Science and Technology)Ó.

\end{document}